\documentclass[12pt]{article}
\usepackage{graphicx}

\newcommand{\e}{{\rm e}}
\newcommand{\ehat}{{\hat{e}}}

\newcommand{\ug}{\; = \;}

\newcommand{\Ebf}{\mbox{\boldmath $E$}}
\newcommand{\Bbf}{\mbox{\boldmath $B$}}
\newcommand{\jbf}{\mbox{\boldmath $j$}}

\newcommand{\Sbf}{\mbox{\boldmath $S$}}
\newcommand{\Abf}{\mbox{\boldmath $A$}}
\newcommand{\rbf}{\mbox{\boldmath $r$}}

\newcommand{\ehatbf}{\mbox{\boldmath $\ehat$}}

\newcommand{\pa}{\partial}

\newcommand{\text}{\rm}

\newcommand{\drm}{{\rm d}}

\newcommand{\infi}{\infty}

\newcommand{\ra}{\rightarrow}

\newcommand{\bb}{\begin{equation}}
\newcommand{\ee}{\end{equation}}
\newcommand{\bega}{\begin{eqarray}}
\newcommand{\ega}{\end{eqnarray}}
\newcommand{\begae}{\begin{eqnarray*}}
\newcommand{\egae}{\end{eqnarray*}}

\newcommand{\h}{\hspace*{4ex}}
\newcommand{\dis}{\displaystyle}

\newcommand{\RSIIrm}{{\rm RS(II)}}

\newcommand{\om}{\omega}

\newcommand{\cent}{\centerline}
\newcommand{\vs}{\vspace*}

\begin{document}

\baselineskip 0.6cm

\begin{center}

{{\large {\bf Cherenkov radiation has nothing to do with X-shaped Localized Waves (Comments on
``Cherenkov-Vavilov Formulation of X-Waves")}}$^{\: (\dag)}$} \footnotetext{$^{\: (\dag)}$ Work
partially supported by FAPESP and CNPq (Brazil), and by INFN, MIUR
(Italy). \ E-mail addresses for contacts: \ recami@mi.infn.it [ER]; \ mzamboni@ufabc.edu.br,
mzamboni@dmo.fee.unicamp.br [MZR]}

\end{center}

\vs{5mm}

\cent{ Michel Zamboni-Rached, }

\vs{2mm}

\centerline{{\em Universidade Federal do ABC, Centro de Ciencias
Naturais e Humanas, Santo Andr\'e, SP, Brazil.}}

\vs{5mm}

\centerline{\rm and}

\vs{3mm}

\cent{ Erasmo Recami }

\vs{2mm}

\cent{{\em Facolt\`a di Ingegneria, Universit\`a statale di Bergamo,
Bergamo, Italy;}}
\cent{{\rm and} {\em INFN---Sezione di Milano, Milan, Italy.}}

\vs{16mm}

{\bf Abstract ---}  {\em The Localized Waves (LW) are
nondiffracting (``soliton-like") solutions to the wave equations,
and are known to exist with subluminal, luminal and superluminal
peak-velocities $V$. \ For mathematical, and experimental,
reasons, the ones that called more attention are the ``X-shaped"
superluminal waves. \ Such waves are associated with a cone, so
that some authors --let us confine ourselves to electromagnetism--
have been tempted to look for links between them and the Cherenkov
radiation: A good example of such attempts is represented by the
article by Walker and Kuperman recently appeared in PRL 99,
no.244802, of Dec.2007. \ However, the X-shaped waves belong to a
very different realm: For instance, they exist even in the vacuum,
independently of any media, as localized non-diffracting pulses
propagating rigidly with a peak-velocity $V>c$, as verified in a
number of papers [cf., e.g., the refs. in the book {\rm Localized
Waves} (J.Wiley; Jan.2008)]. \ We deem it necessary to clarify the
whole question on the basis of a rigorous formalism, in part
original, and of clear physical considerations. In particular we
clarify, by explicit calculations based on Maxwell equations only,
that, at variance with what assumed by some authors: \ (i) the
``X-waves" exist in all space, and in particular inside both the
front and the rear part of their double cone (that has nothing to
do with Cherenkov's); \ (ii) they have not been found
heuristically, via {\rm ad hoc} assumptions, but by use of strict
mathematical (or experimental) procedures; \ (iii) the ideal
X-waves (as well as plane waves) are actually endowed with
infinite energy, but finite-energy X-waves can be easily
constructed (even without space-time truncations): And at the end
of this Ms., by following an original technique, we construct {\rm
exact} finite-energy solutions (totally free from
backward-traveling waves); \ (iv) the large majority of the
researchers working in this area do not aim at using the X-waves
for ``superluminal transmission of information", but are
interested in the circumstance that they are localized waves,
endowed with a self-reconstruction property, and in their
important practical applications (in part already realized, since
1992); \ (v) an actual attempt at comparing Cherenkov radiation
and X-waves would lead one to consider the very different
situation of the (X-shaped, too) field generated by a superluminal
point-charge, a question actually exploited in previous papers,
appeared e.g. in 2004 in Phys.Rev.E \ [Recami, Zamboni-Rached \&
Dartora, PRE 69, no.027602]: We show here explicitly that in this
case the point-charge would not lose energy in the vacuum, and
that the field generated by it would not need to be continuously
feeded by incoming side-waves (as it is indeed
the case for an ideal, ordinary X-wave).} \\


\


\vs{20mm}

The Localized Waves (LW) are nondiffracting (``soliton-like")
solutions to the wave equations, and are known to exist with
subluminal, luminal and superluminal peak-velocities $V$. \ For
mathematical, and experimental, reasons, the ones that called more
attention are the ``X-shaped" superluminal waves. \ Such waves are
associated with a cone, so that some authors ---let us confine
ourselves to electromagnetism--- have been tempted to look for
links between them and the Cherenkov radiation: A good example of
such attempts is represented by the article by Walker and
Kuperman, recently appeared\cite{PRL07}. However, the X-shaped
waves belong to a very different realm: For instance, they exist
even in the vacuum, independently of any media, as localized
non-diffracting pulses propagating rigidly with a peak-velocity
$V>c$, as verified in a number of papers (cf., e.g., the refs. in
the book {\em Localized Waves\/}\cite{book}). \ We deem it
necessary to clarify all the question on the basis of a rigorous
formalism, in part original, and of clear physical considerations.
In particular we clarify, by explicit calculations based on
Maxwell equations only, that, at variance with what assumed by
some authors: \ (i) the ``X-waves" exist in all space, and in
particular inside both the front and the rear part of their double
cone (that has nothing to do with Cherenkov's); \ (ii) they have
not been found heuristically, via {\em ad hoc} assumptions, but by
use of strict mathematical (or experimental) procedures; \ (iii)
the ideal X-waves (as well as plane waves) are actually endowed
with infinite energy, but finite-energy X-waves can be easily
constructed (even without space-time truncations): And at the end
of this work, by following an original technique, we construct
{\em exact} finite-energy solutions (totally free from
backward-traveling waves); \ (iv) the large majority of the
researchers working in this area do not aim at using the X-waves
for ``superluminal transmission of information", but are
interested in the circumstance that they are localized waves,
endowed with a self-reconstruction property, and in their
important practical applications (in part already realized, since
1992); \ (v) an actual attempt at comparing Cherenkov radiation
and X-waves would lead one to consider the very different
situation of the (X-shaped, too) field generated by a superluminal
point-charge, a question actually exploited in previous papers,
appeared e.g. in 2004 in Phys.Rev.E \cite{PRE04}:  We show here
explicitly that in this case the point-charge would not lose
energy in the vacuum, and that the field generated by it would not
need to be continuously feeded by incoming side-waves (as it is
indeed the case for an ideal, ordinary X-wave).

\h As already said, we wish to reply to attempts, like the one in Ref.\cite{PRL07},
at setting forth ``Cherenkov-Vavilov formulations of the X-shaped
Localized Waves" (in the following we shall write only
``Cherenkov" for brevity's sake). The classical problem of the Cherenkov
radiation\cite{FrankNewBook} from a point-charge traveling in a
medium with speed $v$ such that $c_n < v < c$, where $c_n$ and $c$
are the speed of light in the medium and in the vacuum,
respectively, is normally investigated in correct terms. Also in \cite{PRL07} it is presented
in a mathematically correct way; sometimes, however, the language used in such a
context is ambiguous: For example, in \cite{PRL07} the speed $c_n$ in the
medium is just called $c$; furthermore, the point-charge
associated with the Cherenkov radiation is called ``superluminal",
despite the fact that its speed is smaller than the light-speed in
the vacuum. In the existing theoretical and experimental
literature on Localized Waves (see again, e.g., Ref.\cite{book}) and
in particular on X-shaped waves\cite{LZR,LS}, the word
superluminal is reserved to group-speeds actually larger than the
speed of light in the vacuum. These ambiguities create
difficulties, especially whenever
a comparison is made of the Cherenkov radiation with X-shaped
waves. Let us repeat that, in reality, the latter belong to a very different realm; and
exist even in vacuum as localized non-diffracting pulses,
propagating rigidly with superluminal (in our language)
peak-velocities, $V>c$, independently of any media.  Let us try to clarify the whole
subject, and several unjustified implications in papers like \cite{PRL07}, by using
a rigorous formalism and clear physical considerations.

\

\h {\em Our specific considerations and comments are as follows:}

\

{\bf 1) ---} \  As already mentioned, the papers addressing the
Cherenkov effect do incorporate sometimes\cite{PRL07} an ambiguous
terminology (see above), about which the readers must be warned. \
Except for the ambiguous notation, the analysis of the ordinary
scalar-valued Cherenkov radiation in normally correct ---cf.,
e.g., the first part of \cite{PRL07}--- as it can be explicitly
checked; \ anyway, the relevant results are
well-known\cite{FrankNewBook}. \ Except for the ambiguous
notation, the analysis of the ordinary scalar-valued Cherenkov
radiation in the first part of \cite{PRL07} is correct, as it can
be explicitly checked; however, the relevant results seem to be
already known\cite{FrankNewBook}. \ It can be pointed out,
incidentally, that even the vector-valued electromagnetic fields
generated by a {\em really} superluminal point-charge, endowed
with speed $V
> c$, have already been considered and published in \cite{PRE04}
using a procedure totally different, of course, from the one followed e.g.
in \cite{PRL07}.

\

{\bf 2) ---} \  The main equivocal aspect of works like
\cite{PRL07} is that such authors attempt at using the ordinary
Cherenkov radiation to draw conclusions about superluminal
Localized Waves (LW) known as X-shaped waves (or simply X-waves).
The latter, as we said, are nondiffractive solutions to the
homogeneous wave equations and propagate rigidly with superluminal
{\em peak-velocities}. They have been predicted long
ago\cite{Barut,Review86}, have been mathematically
constructed\cite{LZR}, and finite-energy versions of these
wave-packets have been experimentally produced\cite{LS} in the
vacuum, independently of any media. \ We are going to show the
drawbacks of the path followed in articles as \cite{PRL07} by
using explicit calculations.

\h To be clear and self-contained, we initially address the
(simpler) scalar case, in which a field $\psi$ is governed by the
inhomogeneous wave equation

\bb \left( \nabla^2 - \frac{1}{c_n^2}\frac{\pa^2}{\pa t^2}
\right)\psi(\rbf,t) \ug -\frac{4\pi}{c_n}j(\rbf,t)  \label{eo} \ee            

with \ $j = qv \, \delta(\rho) \, \delta(z-vt) / (2\pi\rho)$. \
Here $c_n$ is the speed of light in the considered medium and
$j(\rbf,t)$ is the generating source, assumed to be pointlike and
moving along the positive $z$-axis with subluminal speed $c_n < v
< c$, while $\rho$ denotes the cylindrical radial coordinate.

\h A Green's function for Eq.(\ref{eo}) is given explicitly as

\bb G(\rbf,t,\rbf\,',t') \ug l \; G^+(\rbf,t,\rbf\,',t') + (1-l)
\; G^-(\rbf,t,\rbf\,',t') \label{eq2} \ee                                       

where

 \bb G^{\pm}(\rbf,t,\rbf\,',t') \ug
\frac{\delta\left(t'-(t\mp R/c_n) \right)}{c_nR} \label{gar}\ee                 

and \ $ R \equiv
\sqrt{(z-z')^2+\rho^2+\rho'^2-2\rho\rho'\cos(\theta-\theta')}$.

Quantities $G^+$ and $G^-$ are the retarded and advanced Green's
functions, respectively.

\h When considering only the retarded Green's function ($l=1$),
the solution to the wave equation can be expressed as

\bb \psi(\rbf,t) \ug \int \drm x'^3 \int \drm t' \,
G^+(\rbf,t,\rbf\,',t') \, j(\rbf\,',t') \; . \label{eq4} \ee                   

We can use the expression for $G^+$ given in Eq.(\ref{gar}) for a
direct calculation of the wave function $\psi(\rbf,t)$. However,
for reasons that will be made clear in the sequel, we shall follow
the procedure adopted in papers like \cite{PRL07}. Specifically, we shall
determine the Fourier transform of $G^+$ in the variables $z$ and
$t$, and subsequently the transform of $\psi$ in the same
variables. For $c_n < v < c$, we obtain:

$$ \psi(\rho,\zeta) = \frac{1}{2\pi c_n}
\left[\left(\int_{-\infi}^{0}\drm \om
(-i\pi)qH_0^{2}(\rho\gamma_n^{-1}|\om|/v)e^{i\om\zeta/v} \right) +
\left(\int_{0}^{\infi}\drm \om
(i\pi)qH_0^{1}(\rho\gamma_n^{-1}\om/v)e^{i\om\zeta/v} \right)
\right] $$

where $\zeta \equiv z-vt$, and $H_0^{1,2}$ are the zero-order
Hankel functions of first and second kind. \ Using, next, the
relation $H_0^{2}(x)=-H_0^{1}(-x)$ for $x \geq 0$, the last
equation is reduced to Eq.(7) of Ref.\cite{PRL07}, viz.,

\bb \psi(\rho,\zeta) \ug \frac{1}{2c_n} \int_{-\infi}^{\infi}\drm
\om iqH_0^{1}(\rho\gamma_n^{-1}\om/V)e^{i\om\zeta/V} \label{sh} \ee             

from which one determines the well-known expression for the
Cherenkov radiation

\bb \psi(\rho,z,t) \ug \left\{ \begin{array}{clr}
\dis{\frac{2q\beta_n}{\sqrt{\zeta^2 - \gamma_n^{-2}\rho^2}}} \ \ \
\ \ \
{\rm for}\; & \zeta<-\gamma_n^{-1}\rho  \\
\\
0  \ \ \  & {\rm elsewhere} \; , \end{array}\right. \label{pr}\ee             

where it should be noted that \ $\beta_n \equiv v/c_n$ \ and \
${\gamma_n} \equiv 1 / \sqrt{v^2/c_n^2 - 1}$. \ The radiation
exists only inside the rear part $\zeta = -\gamma_n^{-1}\rho$ of
the Cherenkov cone.

\h As to the vectorial case, physically more significant, let us
only mention that is can be constructed by adopting the Lorentz
gauge, and by considering the vector potential $\Abf =
A_z\ehatbf_z$, with $A_z \equiv \psi$ and the current density
$\jbf = j_z\ehatbf_z$, with $j_z \equiv j$.

\

\h {\bf 2a) -- } \ Some authors, as the ones of Ref.\cite{PRL07},
have been induced at looking for a connection between the
Cherenkov emission and X-shaped waves. \ The simple X-waves to
which many authors refer to are solutions to the homogeneous
scalar wave equation: They are wave functions of the type
$X(\rho,z,t)=X(\rho,\zeta)$, with $\zeta \equiv z-Vt$ and
$c_n<V<\infty$, and can be obtained by suitable superpositions of
axially symmetric Bessel beams, propagating in the positive
$z$-direction with the same phase-velocity (cf., e.g.,
Refs.\cite{LZR}); precisely,

\bb \psi_{\rm X}(\rho,\zeta) \ug \int_{0}^{\infty}\drm\,\om S(\om)
J_0(\rho\frac{\om}{V}\sqrt{V^2/c^2-1})e^{i\om/V\zeta} \label{sb}
\ee                                                                         

where $J_0(.)$ is an ordinary zero-order Bessel function and
$S(\om)$ is the temporal frequency spectrum. \ Incidentally, let
us stress that Eq.(\ref{sb}) represents a particular case of the
superluminal waves, which in their turn are just a particular case
of the (subluminal, luminal or superluminal) Localized Waves. For
the specific spectrum $S(\om) = \exp [-a\om]$, with $a$ a positive
constant, one obtains the zero-order (classic) X-wave:

\bb X \equiv X(\rho,\zeta) \ug \frac{V}{\sqrt{(aV-i\zeta)^2 +
(V^2/c^2
-1)\rho^2}} \; . \label{x} \ee                                               

Authors like the ones in Ref.\cite{PRL07} are led to attempt a comparison
between the Cherenkov effect and the zero-order X-wave by the
apparent mathematical similarity of equations (\ref{sh}) and
(\ref{sb}) [which correspond, for instance, to equations (7) and (13) of
Ref.\cite{PRL07}.] \ To be more specific: \ (i) we have seen that
for the inhomogeneous wave-equation (\ref{eo}), a Cherenkov
solution of the type given in Eq.(\ref{pr}) exists only inside the
cone rear part $\zeta = -\gamma^{-1}\rho$. \ To obtain a solution
existing only inside the cone forward part $\zeta =
\gamma^{-1}\rho$, one should make use of the advanced Green's
function $G^-$: This ---to go on with our example--- induced the authors in \cite{PRL07}
to state that the forward part of the X-wave is non-causal; \ (ii) Another
point that misled Walker and Kuperman\cite{PRL07} is that, if one
puts $S(\om)=i$ into the $\psi_{\rm X}$-wave (\ref{sb}), that is,
if one sets $a=0$ in Eq.(\ref{x}) and multiplies it by $i$, one
obtains

\bb \widetilde{X}(\rho,\zeta) \ug \frac{V}{\sqrt{\zeta^2 -
(V^2/c^2 -1)\rho^2}} \; , \label{x2} \ee                                      

which is mathematically identical, apart from a constant, to the
Cherenkov solution in Eq.(\ref{pr}]), with a real part existing
this time inside both the rear cone $\zeta = -\gamma^{-1}\rho$ and
the forward cone $\zeta = \gamma^{-1}\rho$. \ The statement in
papers like \cite{PRL07} that the advanced part of the zero-order X-wave [cf.
Eq.(\ref{x})] is non-causal is due to an illicit extrapolation.
Being a solutions to the homogeneous wave equation, the X-wave
cannot admit singularities. In contrast, the solution given in
Eq.(\ref{x2}), which was obtained from the Bessel beam
superposition (\ref{sb}) using a {\em constant} spectrum
$S(\omega)$, does have singularities and cannot be considered a
solution to the homogeneous wave equation. Actually, the real part
of the solution (\ref{x2}) can be an acceptable solution for the
inhomogeneous case only: Indeed, we shall show later on that it
represents the field of a point-charge traveling with speed $V>c$
when using as a Green's function the expression $G = G^+/2 +
G^-/2$, half retarded and half advanced (which means, again, that
it refers to an inhomogeneous problem).

\h It should be pointed out that the Bessel beam synthesis in
Eq.(\ref{sb}) is a particular case of more general spectral
representations leading not only to infinite but also to finite
energy superluminal LWs\cite{book,LZR}.  Not less important, even
a finite-energy close replica of the the classic X-wave (\ref{x})
(endowed, incidentally, also with a finite field-depth) can be
generated in a {\em causal} manner by means of a ``dynamic" finite
aperture (antennas, holographic or optical elements, etc.). For
instance, it is enough an array of circular elements excited
according to the function $X(\rho,z=0,t)$, given by equations
(\ref{sb}) or (\ref{x}) on the aperture plane located at $z=0$.
The emitted finite-energy X-wave can be calculated by means of the
Rayleigh-Sommerfeld (II) formula \cite{Goodman}

\bb \psi_{\RSIIrm}
\left(\rho,z,t\right)=\int_0^{2\pi}\drm\phi'\int_0^{D/2}
\drm\rho'\rho' \frac{1}{2\pi R}\left\{[X]\frac{(z-z')}{R^2} \, +
\, [\partial_{c_n t'}X]\frac{(z-z')}{R}\right\} \; . \label{rs}\ee             

The quantities inside the square brackets are evaluated at the
retarded time $c_n t'=c_n t-R$. The distance
$R=\sqrt{\left(z-z'\right)^2+\rho^2+\rho'^2-2\rho\rho'\cos\left(\phi-\phi'%
\right)}$ is now the separation between source and observation
points, and the aperture diameter is denoted by $D$. \ The depth
of field of the particular solution given in Eq.(\ref{rs}) is
known\cite{JLZ} to be $Z=R \gamma_n$.

\h What stated above has been theoretically (even via numerical
simulations...) and experimentally verified, as published in a large
number of well-known papers: see, for example,
Refs.\cite{LZR,LS,SB,JLZ,MRH,IEEE,book}.

\

\h {\bf 2b) --} \ We agree with authors as Walker and
Kuperman\cite{PRL07} that the superluminal spot of any X-wave is
fed by the waves coming from the elements of the aperture, and
that these waves carry energy with at most luminal ($V=c_n$)
speed. In such cases, the X-wave {\em intensity peaks} at two
different locations are not causally correlated. But such a
correct claim, contained also in \cite{PRL07}, is well known and
firmly accepted since the nineties by practically all the scholars
working in the area of LWs (cf., e.g., the references in
\cite{Introd} and \cite{IEEE}, as well as
Refs.\cite{book,LZR,JLZ,SB}): The claim in \cite{PRL07} that
efforts in the area of X-waves are aimed at transmitting
information superluminally is not correct. The large majority of
the experts in such a subject are interested in X-waves due to
their spatio-temporal localization, unidirectionality,
``soliton-like", and self-reconstruction properties in the
near-to-far zone. Such properties bear interesting consequences,
from theoretical and experimental points of view in all sectors of
physics in which a role is played by a wave equation (including
---{\em mutatis mutandis\/}--- elementary particle physics, and
even gravitation).

\

{\bf 3) ---} \ The statement, contained in papers like
\cite{PRL07}, that the (ideal) zero-order X-wave in Eq.(\ref{sb})
has infinite energy (as well as the plane-waves) does not convey
new information. The fact that such a solution needs to be fed for
an infinite time has been known since the start of LW theory. \ In
any case, as mentioned earlier, such a problem can be overcome
either by using (cf. Eq.({\ref{rs})) apertures finite in space and
time, i.e., by truncating the X-wave, or by constructing exact,
analytical finite-energy solutions\cite{MRH,SB}. \ For reasons of
space, we shall show {\em only briefly}, but in an original
rigorous way, how closed-form solutions of the latter type can be
actually constructed, without any recourse to the
backward-traveling waves that trouble the ordinary approaches: See
point {\bf 5)} below.

\h Here, for the moment, let us recall that X-waves endowed by themselves
with finite energy even without truncation have been easily constructed
in the past by use of diffraction integrals: See, for instance,
the approximate solution in Eqs.(2.31),(2.32) of Ref.\cite{MRHbook}, that is,
the ``SMPS pulse", which is depicted in Figs.1. A finite-energy X-wave gets
deformed while propagating: and Fig.1(b) shows the pulse in
Fig.1(a) after it has travelled 50 km.

\begin{figure}[!h]
\begin{center}
 \scalebox{1.0}{\includegraphics{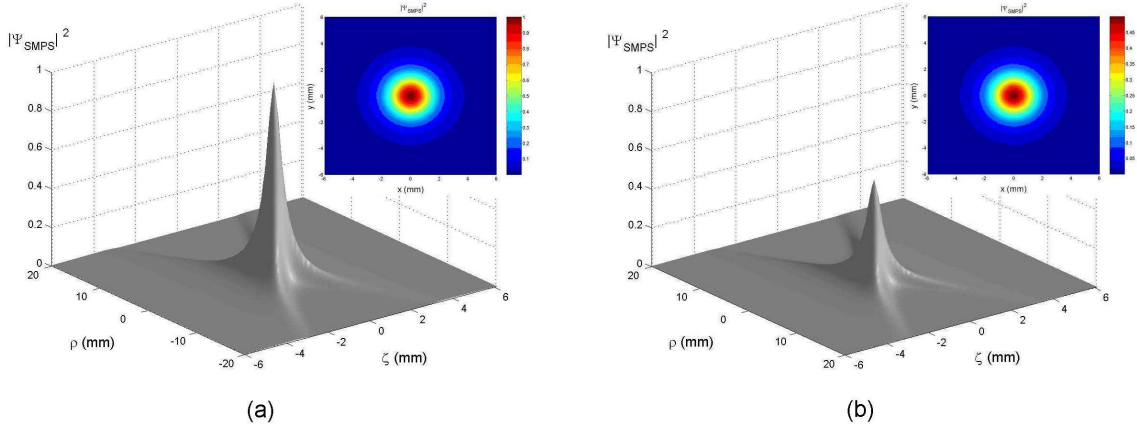}}
\end{center}
\caption{Example of an X-type LW endowed with finite energy (even
without truncation), and that consequently gets deformed while
propagating: (b) represents the pulse depicted in (a) after it has
traveled 50 km. See Ref.\cite{MRHbook}, and the text.}
 \label{fig1}
\end{figure}

\h Let us emphasize, as well, that the formulations leading to
LWs, and to X-waves, are not chosen {\em ad hoc}, as believed in
papers as \cite{PRL07}, but are to be based on proper choices of the spectra
(which imply a specific space-time coupling\cite{MRH,SB,theses})
and of the Bessel functions (in order to avoid singularities both
at $\rho=0$ and at $\rho=\infty$): In contrast, the choice
suggested, e.g., in Eq.(15) of \cite{PRL07} presents singularities. \ To
be clearer, let us observe that a general solution to the scalar
homogeneous wave-equation in free space can be written [when
eliminating evanescent waves] in the form

\bb \psi(\rho,\phi,z,t) \ug
\sum_{\nu=-\infty}^{\infty}\left[\int_{0}^{\infty}\drm
\om\,\int_{-\om/c}^{\om/c}\drm k_z A_\nu(k_z,\om)
J_\nu\left(\rho\sqrt{\frac{\om^2}{c^2} - k_z^2} \right)e^{ik_z
z}e^{-i\om t}e^{i \nu \phi} \right] \label{S2geral2} \ee                 

by considering positive angular frequencies $\om$ only. \ For
obtaining ideal LWs propagating along the positive $z$-direction
with peak-velocity $V$ (that can assume a priori any value $0\leq
V \leq \infty$), the spectra $A_\nu$ {\em must} have the
form:\cite{MRH,book,theses}

\bb A_\nu(k_z,\om) \ug \sum_{\mu=-\infty}^{\infty}\,S_{\nu
\mu}(\om) \; \delta [\om - (Vk_z + b_\mu)] \label{S2speclw} \ee          

where \ $b_\mu = 2\pi\mu V/\Delta z_0$, \ and it can be easily
shown\cite{MRH,theses} that solution (\ref{S2geral2}) possesses
the important property \ $\psi(\rho,\phi,z,t) = \psi(\rho,\phi,z +
\Delta z_0,t + \Delta z_0/V)$, \ where $\Delta z_0$ is a chosen
space-interval along $z$. \ The last two equations already show
that LWs are not to be found by ``ad hoc" assumptions: Equation
(\ref{S2speclw}) does explicitly show that
---as mentioned above--- the ideal LWs exist only in
correspondence with linear relations between $\om$ and $k_z$, that
is, with {\em specific} space-time couplings. \ By assuming in
particular \ $A_0(k_z,\om) \equiv A_\nu(k_z,\om) = \delta_{\nu 0}
\; S(\om) \; \Theta(k_z) \; \delta [\om - (Vk_z + \alpha_0)]$, \
where the Heaviside function $\Theta$ does eliminate, as desired,
any backward components, and $\alpha_0$ is a constant, we can
construct an infinite number of subluminal, luminal, or
superluminal Localized Solutions, with axial symmetry, and in
closed form. For instance, the X-shaped waves represented in
Eq.(\ref{sb}), and used for the sake of comparison also in
Ref.\cite{PRL07}, correspond to the particular value $\alpha_0 =
0$. [To get finite-energy solutions one has to abandon the strict
request of a linear relation between $\om$ and $k_z$, and impose
instead that the spectral functions $A(k_z,\om)$ possess non
negligible values only {\em in the vicinity} of a straight line of
the mentioned type\cite{MRH}: This is briefly exploited under
point {\bf 5)} below].

\

{\bf 4) ---} \ We have already established that the analogy
attempted in works like \cite{PRL07} between the Cherenkov
radiation and the X-wave solution is not justified even in
material media. \ In the case of vacuum, the aforementioned
analogy should have rather led one to consider the field generated
by a {\em really} superluminal point-charge: A problem that was
investigated in \cite{PRE04}, and refs. therein. \ In such a
situation, the point-charge superluminally traveling in the vacuum
is not expected to radiate due to physical reasons published long
ago\cite{LNC73,Review86,Review74}. Such a charge does not radiate
in its rest-frame\cite{Review86,Review74} and, consequently, does
not radiate also according to observers for whom it is
superluminal\cite{LNC73}. \ We shall establish this result below,
by explicit calculations based on Maxwell equations only.

\h Let us consider the wave equation (\ref{eo}) in vacuum ($c_n
\ra c\/$) with a superluminally moving ($v \ra V>c\/$) point
charge source. Use of the Green's function $G = (G^+ + G^-)/2$
(cf., e.g., Refs.\cite{SS,Saari2}) yields the following integral
representation for the solution:

\bb \psi(\rho,\zeta,t) \ug \frac{q}{c} \int_{0}^{\infi}\drm \om \;
N_0(\rho {\om \over V} \gamma^{-1}) \; \cos({\om \over V} \zeta)
\; , \label{eq12} \ee                                                            

where $N_0$ is the zero-order Neumann function, and, now,
$\beta_n$ and $\gamma_n$ have been replaced with $\beta \equiv \
V/c$ and $\gamma^{-1} \equiv \sqrt{V^2/c^2 - 1}$. \ The
integration can be carried out explicitly, and yields, for the
field generated by a point-charge traveling superluminally in the
vacuum, the expression

\bb \psi(\rho,z,t) \ug \left\{ \begin{array}{clr} \dis{q \beta
\left[ \zeta^2 - \rho^2 \gamma^{-2} \right]^{-1/2}} \ \ \ \ \ \ \
\ \ \
& {\rm when} \ \ \ \ \ \ \ \ \ \ 0 < \rho\gamma^{-1} < |\zeta|  \\
\\
0  \ \ \ \ \ \ \ \ \ \ & {\rm elsewhere}  \; .
\end{array}\right. \label{(*)} \ee                                              

This solution is different from zero inside the rear {\em and}
front parts of the unlimited double cone\cite{PRE04,Introd}
generated by the rotation around the $z$-axis of the straight
lines $\rho = \pm \gamma \zeta$, in agreement with the predictions
of the ``extended" (or, rather, ``non-restricted") theory of
Special Relativity\cite{Review86,Review74}. \ The expression in
Eq.(\ref{(*)}) is precisely equivalent to the solution given by
Eq.(8) in our Ref.\cite{PRE04}; except for a constant that was
wrong therein.

\h Going on to the (physically more suited) vectorial formalism,
adopting the Lorentz gauge, and choosing a current density \ $\jbf
= j_z\ehatbf_z$; \ $j_z \equiv j$, \ a scalar electric potential
$\phi = c \psi / V$ \ and a vector magnetic potential $A \equiv
\psi\ehatbf_z$ \ (cf. Fig.2 in Ref.\cite{PRE04}, which refers to a
negative point-charge), one obtains, in analogy to \cite{PRE04},
the electric and magnetic fields

\bb \Ebf(\rho,\zeta) = -q\gamma^{-2} Y (\rho \, \ehatbf_\rho +
\zeta \ehatbf_z)\, ; \ \ \ \ \ \ \Bbf = -q\beta\gamma^{-2} Y
\rho \, \ehatbf_\theta \; , \label{EB} \ee                                     

in Gaussian units; where

$$Y \; \equiv \; \left[ \zeta^2 - \rho^2 \gamma^{-2} \right]^{-3/2} $$

inside the double cone (i.e., for $0<\rho\gamma^{-1}<|\zeta|$),
while $Y = 0$ outside it (that is, $\Ebf$ and $\Bbf$ are zero
outside it). \ The corresponding Poynting vector is given by

\bb \Sbf = {c \over {4\pi}} q^2 \beta \gamma^{-4} Y^2 \rho \,
(\rho \, \ehatbf_z - \zeta \ehatbf_\rho) \; . \label{eq15} \ee                    

The total flux, through any closed surface containing the
point-charge at the considered instant of time, is equal to zero.
Thus, a point-charge traveling at a constant superluminal speed in
vacuum does not radiate energy. This fact is depicted and
explained in an intuitive way in Fig.4 of Ref.\cite{PRE04}, which
originally appeared in Refs.\cite{Barut,Review86} and, for
clarity, is reproduced in this article as Fig.2.

\begin{figure}[!h]
\begin{center}
 \scalebox{1.4}{\includegraphics{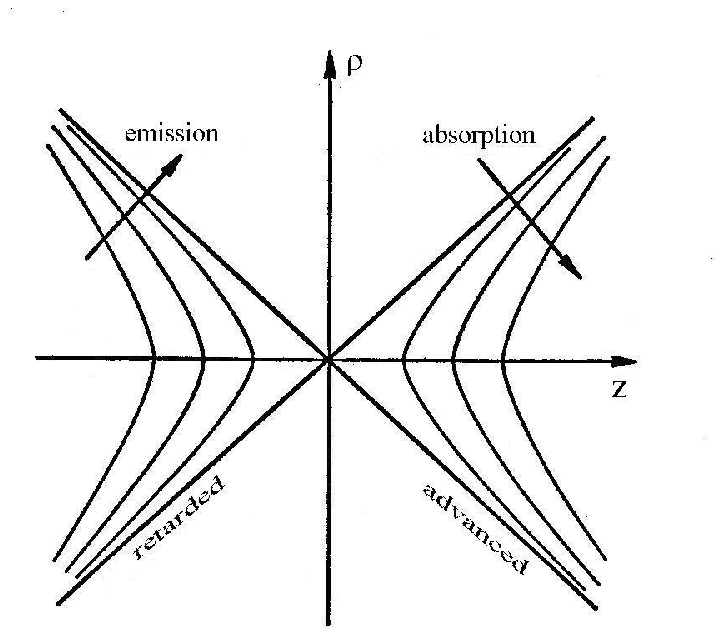}}
\end{center}
\caption{This figure ---appeared in Ref.\cite{PRE04}, but taken
from Ref.\cite{Review86}--- does intuitively show, among the
others, that a Superluminal
charge\cite{LNC73,Review74,Barut,Review86,IEEE,Introd} traveling
at constant speed in the vacuum, would {\em not} lose energy: see
\cite{PRE04} and the text.}
 \label{fig2}
\end{figure}

\h We wish to emphasize that in the present case the field needs
not be fed, at variance with the case of the ordinary X-waves.
Once more, one can see that the analogy exploited in papers like \cite{PRL07}
between the Cherenkov effect and the X-waves completely breaks
down in the case of the vacuum.

\h Eventually, most authors do not address many of the other
interesting physical points. For example, no mention
appears in \cite{PRL07} of the fact that a superluminal charge is expected to
behave as a magnetic pole, in the sense fully clarified in
Refs.\cite{PLB,Barut,Review86}. One can see even from
Eqs.(\ref{EB}) that $\Ebf \ra 0$, and one is left with a pure
magnetic field, in the limit $V \ra \infty$.

\

{\bf 5) ---} \ {\em At last,} as anticipated in point {\bf 3)}
above, finite-energy solutions can be obtained in closed form
without any recourse to the backward-traveling waves that trouble
the usual approaches (even if the intervention of such components
has been already minimized in Refs.\cite{MRH,SB}, at the cost
---however--- of going on to frequency spectra with a very large
bandwidth).  \ In fact, when confining ourselves to superluminal LWs
with axial symmetry, let us put in Eq.(\ref{S2geral2}) \
$A_\nu(k_z,\om) = \delta_{\nu 0} \, A(k_z,\om)$, \ and
adopt the ``unidirectional decomposition"

$$ \zeta \equiv z - Vt \, ; \ \ \ \ \ \ \ \ \ \ \eta \equiv z -ct \; .$$

In terms of the new variables\footnote{The same variables were
adopted in Ref.\cite{BS} in the paraxial approximation context,
while we are addressing the general exact case} and confining
ourselves now to $V>c$, equation (\ref{S2geral2}) can be rewritten
as

$$\psi(\rho,\zeta,\eta) = (V-c) \int_0^\infty \drm \sigma
\int_{-\infty}^\sigma \drm \alpha  J_0 \left( \rho
\sqrt{\gamma^{-2} \sigma^2 - 2(\beta - 1) \sigma \alpha} \right)
\exp [-i\alpha \eta]  \exp [i \sigma \zeta] A(\alpha,\sigma)$$

where \ $\alpha \equiv (\om - Vk_z)/(V-c)$ \ and \ $\sigma \equiv
(\om - ck_z)/(V-c)$.

\begin{figure}[!h]
\begin{center}
 \scalebox{1.0}{\includegraphics{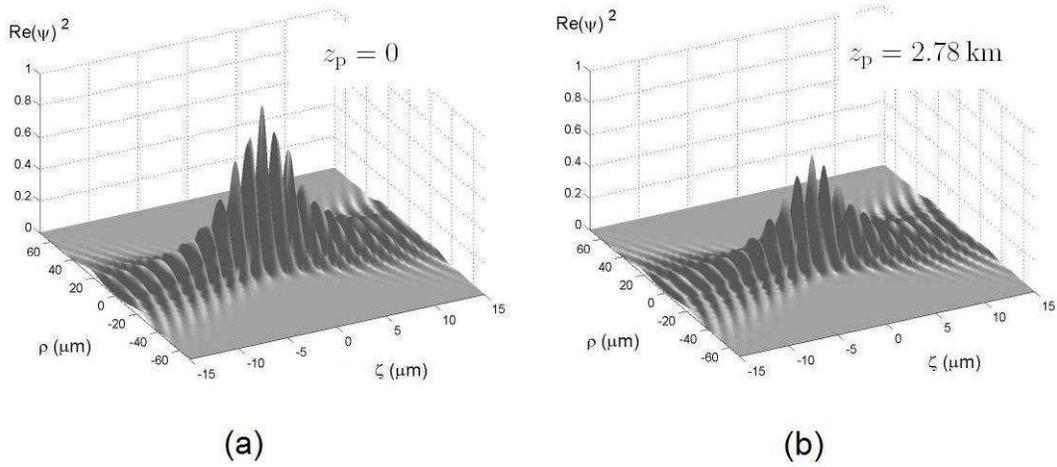}}
\end{center}
\caption{Example of a finite-energy X-type LW, corresponding to an
exact, analytic solution of Eq.\ref{eq18}, totally free from
backward-components. This figure represents the real part of the
field, normalized at $\rho=z=0$ for $t=0$, with the choices
$a=3.99\times10^{-6}\,$m, \ $d=20\,$m, \ $V=1.005\,c$, and
$\alpha_0=1.26\times 10^7\,{\rm m}^{-1}$. In this case the
frequency spectrum starts at $\om \equiv \om_{\rm min} \approx
3.77\times 10^{15}\,$Hz and afterwards decays exponentially with
the bandwidth $\Delta\om \approx 7.54\times 10^{13}\,$Hz. The
value $\om_{\rm min}$ can be regarded as the pulse central
frequency; since $\Delta\om / \om_{\rm min} << 1$, it exists a
well-defined carrier wave, which does clearly show up in the
plots. \ Any finite-energy LW gets deformed while propagating, and
(b) represents the pulse depicted in (a) after it has traveled
2.78 km.}
 \label{fig3}
\end{figure}

\h As mentioned above, ideal (infinite energy) superluminal LWs
are got by imposing the linear constraint \ $A(\alpha,\sigma) =
B(\sigma) \, \delta (\alpha + \alpha_0)$. \ By contrast,
finite-energy superluminal LWs are obtained by concentrating the
spectrum $A(\alpha,\sigma)$ in the vicinity of the straight line
$\alpha = -\alpha_0$. \ By choosing for example

\bb A(\alpha,\sigma) \ug {{\Theta(-\alpha - \alpha_0)} \over
{V-c}} \; \e^{d\alpha} \; \e^{-a\sigma} \; , \label{eq17} \ee                    

we get the {\em finite-energy exact solutions} (free from any
backward-components):

\bb \psi(\rho,\zeta,\eta) = {X \over V Z} \; \e^{-\alpha_0 Z} \; ,
\label{eq18} \ee                                                                  

where $X$ is defined in eq.(\ref{x}), quantities $a$, $d$ and
$\alpha_0$ are positive constants, and

$$ Z \, \equiv \, (d-i\eta) - {c \over {V+c}} \, (a-i\zeta
-VX^{-1}) \; . $$

\h In Figs.3 we show one of such finite-energy Superluminal LWs,
corresponding to an exact, analytic solution totally free from
backward-travelling components. As we know, any finite-energy X-wave
gets deformed while propagating: and Fig.3(b) shows the pulse in
Fig.3(a) after it has travelled 2.78 km.

\

\

{\bf 6) \ Acknowledgements} --- The authors are greatly indebted
to I.M.Besieris for having called their attention to
Ref.\cite{PRL07} and for a very careful, extensive and painstaking
revision of this work; and to I.M.Besieris, P.Saari, and
C.A.Dartora for many stimulating discussions.

\end{document}